\input{epsf}
\documentstyle[12pt]{article}
\newcommand{\oas}{$O(\alpha_s)$\ }
\newcommand{\oass}{$O(\alpha_s^2)$\ }
\newcommand{\bxee}{B \to X_s e^+e^-}
\newcommand{\bc}{\begin{center}}
\newcommand{\ec}{\end{center}}
\newcommand{\ve}{\vspace}

\newcommand{\beq}{\begin{equation}}
\newcommand{\eeq}{\end{equation}}
\newcommand{\bey}{\begin{eqnarray}}
\newcommand{\eey}{\end{eqnarray}}
\newcommand{\bdm}{\begin{displaymath}}
\newcommand{\edm}{\end{displaymath}}
\newcommand{\hs}{\hat{s}}
\newcommand{\as}{\frac{\alpha_s}{4 \pi}}
\newcommand{\bxg}{B \to X_s \gamma}
\newcommand{\nn}{\nonumber}
\begin{document}
\begin{titlepage}
\setcounter{page}{0}
\rightline{Preprint YerPhI 1572(9)-2001}
\ve{2cm}
\bc
{\large {\bf \oas Corrections to $\bxee$ Decay in the 2HDM.}}
\ec
\vspace{0.2cm}
\bc
G. K. Yeghiyan \\
{\it Yerevan Physics Institute, Yerevan, Armenia.}
\ec
\ve{1cm}
\centerline{\bf Abstract}
\vspace{0.1cm}
\oas QCD corrections to the inclusive $\bxee$ decay are investigated within the
two - Higgs doublet extension of the standard model (2HDM). The analysis is
performed in the so - called off -resonance region; the dependence of the
obtained results on the choice of the renormalization scale is examined in
details. It is shown that 
\oas corrections can suppress the $\bxee$ decay width up to $1.5 \div 3$ times
(depending on the choice of the dilepton invariant mass $s$ and the low -
energy
scale $\mu$). As a result, in the experimentally allowed range of the 
parameter
space, the relations between the $\bxee$ branching ratio and the new physics 
parameters
are strongly affected. It is found also that though the renormalization scale
dependence of the $\bxee$ branching is significantly reduced, 
higher order 
effects in the perturbation theory can still be nonnegligible. 
\end{titlepage}
\newpage
1. Rare B - meson decays can serve as an important source of information on new
physics beyond the standard model. The first experimental evidence for these
decays has been observed by CLEO \cite{1} for the exclusive $B \to K^* \gamma$
channel. Later on the branching ratio of the inclusive $\bxg$ decay has been
measured by CLEO, ALEPH and Belle collaborations \cite{2}-\cite{4}. The weighted
average for this decay branching is \cite{5}
\beq
B^{exp}(\bxg) = (2.96 \pm 0.35) \times 10^{-4} 
\eeq
Recently the evidence for the rare exclusive channel 
$B \to K \mu^+ \mu^-$ has been also observed \cite{26}.

The experimental result for the $\bxg$ branching is in a good agreement with 
the SM predictions (see \cite{5} and
references therein). The new physics contribution to this decay width must be
small enough to avoid the contradiction with the experiment, hence studying
$\bxg$ decay one can get some constraints on the new physics parameters.

The another popular inclusive rare B decay mode, $\bxee$, has not been 
observed yet (only upper bound on its branching ratio exists \cite{26}),
however it is expected to be measured during the forthcoming experiments at the
B - factories. 
Then, analogously to $\bxg$, the study of $\bxee$ can
provide 
some information on the physics, which occur above the scale $\sim 100$GeV.
\\
\\
2. The $\bxee$ decay has been studied within the standard model and it
extensions in \cite{6}-\cite{8} and \cite{9}-\cite{11} respectively. In the
latter works it has been shown that new physics contribution can make $\bxee$
branching ratio two and more times larger than in the SM. These
calculations have been performed in the next-to-leading order (NLO) of the
perturbation theory, which includes $O(\alpha_s^{-1})$ and $O(1)$ contribution
to $\bxee$ decay. On the other hand, it has been proven in \cite{8} that
to this order the obtained results may suffer from the uncertainties, connected
with those in the choice of the low-energy scale $\mu \sim m_b$ and the heavy 
mass (matching) scale $\mu_W \sim M_W,$ $m_t$.

Recently the \oas corrections to the $\bxee$ decay branching have been 
calculated
in \cite{12,13}. The calculations have been performed in the off-resonance
region, corresponding to $0.05 < \hs < 0.25$, where $\hs$ is the dilepton
invariant mass, normalized over the b - quark mass.
It has been found that within the standard model these
corrections reduce the above-mentioned uncertainties about two times.

The aim of the present paper is studying the impact of the \oas corrections
on the $\bxee$ decay rate in the two-Higgs doublet extension of the 
standard model
(2HDM). The investigation is carried out for the most general version of the 
2HDM
(so-called Model~III), where sizable deviations from the SM result are possible.
The deviations from the SM predictions (called new physics effects) occur
due to diagrams with the charged Higgs boson mediated loops. It is shown
that when the contribution of these diagrams is sizable, \oas contribution
suppresses the $\bxee$ decay width up to $1.5 \div 3$ times, depending on the
choice of the dilepton invariant mass and the low - energy scale. As a result,
in the experimentally allowed region of the parameter space
the behavior of the $\bxee$ branching ratio with the new physics parameters is
changed drastically. It is also pointed out that the dependence of $B(\bxee)$ on
the parameters of theory can further be modified. While due to
the \oas corrections low - energy scale dependence of the obtained 
results is
very small (or even negligible), the matching scale dependence remains large
enough. This indicates that higher order effects in the perturbation theory can
be numerically relevant as well. \\
\\
3. It is known that in the SM the decay $\bxee$ is loop-induced:
in the lowest order it proceeds via exchange of the up-type quarks
and $W^{\pm}$ boson in the loops. In the 2HDM there are
additional diagrams with W-boson replaced by the charged
Higgs boson ($H^{\pm}$). The interaction of the charged Higgs boson with 
quarks
may be written in the following form \cite{14}:
\beq
\frac{g}{\sqrt{2}M_W} \left(\xi_t m_t H^- (\bar{d},\bar{s},\bar{b})_L 
\left(\begin{array}{c} V^*_{td}\\ V^*_{ts}\\ V^*_{tb}\end{array}\right) 
t_R  + \xi_b m_b H^+(\bar{u},\bar{c},\bar{t})_L 
\left(\begin{array}{c} V_{td}\\ V_{ts}\\ V_{tb}\end{array}\right) 
b_R + h.c \right) 
\eeq
where $V$ is the CKM matrix, $g$ is the weak coupling constant, $m_t$, $m_b$ are the running top and
bottom masses (for the relation between the pole and running quark masses see
e.g. \cite{15}), and the parameters $\xi_t$ and $\xi_b$ are the functions of
the Higgs doublet vev's, top and bottom masses and the couplings of Yukawa 
interaction of t- and b- quarks with the Higgs doublets \cite{16}. 
Due to the Higgs doublet vacuum phase, $\xi_t$ and $\xi_b$ are complex
in general. Notice also that while $|\xi_t| \sim 1$ or smaller, 
$|\xi_b|$ can be much larger
than unity, unless it contradicts with the experimental constraints on the 
$\bxg$ branching.

To avoid flavor changing neutral currents (FCNC) in the Lagrangian, one usually
considers the simplified versions of the 2HDM: Model~I, where only one of 
the Higgs
doublets interacts with quarks and the Model~II, where one of the Higgs doublets
interacts with the up-type quarks and the second one does with the down-type
quarks. In these versions of the 2HDM, $\xi_t=\xi_b=-\cot{\beta}$ and
$\xi_t=-\cot{\beta}$, $\xi_b=\tan{\beta}$ respectively ($\tan{\beta}$ is the
Higgs vev's ratio). Notice however that due to the stringent constraints on the 
new physics parameters, in the
Models~I~and~II the predictions for the $\bxee$ decay branching coincide 
(with $(10 \div 15)\%$ accuracy) to those of the standard model.

In the present paper the most general version of the 2HDM (Model~III), 
where both of
the Higgs doublets interact with both the up-type and down-type quarks, is
considered (neglecting possible FCNC's in the Lagrangian). In this case, due to
larger parameter space, $\bxee$ branching can be up to three times larger than
in the SM. It is worth also to mention that the results derived for the
Model~III may be also considered valid for multi-Higgs doublet models with only
one light charged Higgs boson. \\
\\
4. The  $B \to X_s e^+ e^-$ decay is studied, 
using the effective theory with five quarks obtained by integrating out
the heavy degrees of freedom which are
W and Z bosons, t-quark and the charged Higgs boson.
The effective Hamiltonian for  the decay $\bxee$ can be
written as
\begin{eqnarray}
H_{eff}\Big(b \to s e^+ e^- (+g)\Big)=
-\frac{4G_F}{\sqrt{2}} \left [\lambda_{t}^{s}
\sum_{i=1}^{10}C_i(\mu) O_i(\mu) 
-\lambda_u^{s}\sum_{i=1}^{2}C_i(\mu) \Big(O_{i}^u(\mu)
- O_i(\mu)\Big) \right]    
\end{eqnarray}
where $\lambda_t^{s}=V_{ts}^* V_{tb}$,
$\lambda_u^{s}=V_{us}^* V_{ub}$, $C_i$ are the coefficients of the Wilson 
expansion and the full set of operators $O_i$ can be found elsewhere 
\cite{12,13}. As it was mentioned above, current study includes only so-called 
off-resonance region, where $0.05 < \hs < 0.25$. In this case $\bxee$
decay is well described by $b \to s e^+ e^-$ and $b \to s e^+ e^- g$ partonic
transitions. The
calculation of these partonic transitions includes the following three steps:
\begin{enumerate}
\item{The Wilson coefficients $C_i$ at the heavy mass scale, $\mu_W \sim M_W$,
$m_t$, must be calculated, matching the effective and full theories. In the
next-to-next-to-leading order (NNLO) the matching has to be done at the 
$O(\alpha_s)$ level, i.e. 
$C_i(\mu_W)=C_i^{(0)}(\mu_W)+\alpha_s/(4\pi) C_i^{(1)}(\mu_W)$. When determining
matching conditions for the Wilson coefficients one must, generally speaking,
take into account the difference between the electroweak breaking scale and new
physics scale. In the 2HDM, Model~III, such a problem occurs when 
$m_{H^+} \gg \mu_W$. Such values of the charged Higgs mass are out of the scope
of this paper: here one takes $m_{H^+}=100$GeV, 200GeV and 400GeV.}
\item{The renormalization group equations (RGE) must be used to obtain the
Wilson coefficients at the low-energy scale $\mu \sim m_b$. In the
next-to-next-to-leading order this step requires
the knowledge of the anomalous dimension matrix up to the order 
$\alpha_s^2$.}
\item{The matrix elements of the operators $O_i$ for the processes
$b \to s e^+ e^-$, $b \to s e^+ e^- g$ have to be calculated.}
\end{enumerate}
Using this procedure, one finds that in the standard model the differential
branching ratio of $\bxee$ decay is given by the following expression \cite{13}:
\begin{eqnarray}
\nonumber
\frac{dB(B \to X_s e^+ e^-)}{d\hat{s}}&=&\frac{\alpha_{em}^2}{4\pi^2}
\frac{|\lambda_t^s|^2}{|V_{cb}|^2|}\frac{(1 -\hs)^2}{g(z) k(z)} \times
\Biggl((1+ 2 \hs) \left(|C_{9,new}^{eff}(\mu)|^2 + 
|C_{10,new}^{eff}(\mu)|^2\right) +\\ 
&&4\left(1 + \frac{2}{s}\right) |C_{7,new}^{eff}(\mu)|^2 + 
12 Re\left(C_{9,new}^{eff*}(\mu) C_{7,new}^{eff}(\mu) \right)\Biggr) 
B_{sl} 
\end{eqnarray}
where $\alpha_{em}$ is the electromagnetic coupling constant, $B_{sl}$ is the
$B \to X_c e^+ \nu$ (experimental) branching ratio, z is the ratio of c- and b-
quark masses squared and the functions g(z) and k(z) are given in \cite{12}. The
{\it effective new} Wilson coefficients are related with the {\it old} ones,
given e.g. in \cite{7,12,17}, as\footnote{Our notation for the Wilson
coefficients is different from that in ref. \cite{12}: here the superscript
$^{(0)}$ denotes $O(\alpha_s^{-1})$ and $O(1)$ contribution to $C_i$, whereas 
$^{(1)}$ always denotes \oas contribution.} 
\begin{eqnarray}
\nn
&& C_{9,new}^{eff}= C_{9}^{(0) eff} + \as C_{9}^{(1) eff} + 
\as \left(4 C_{9}^{(0) eff} \omega_9(\hs) - C_{1}^{(0)} F_1^{(9)} -
C_{2}^{(0)} F_2^{(9)} - C_{8}^{(0) eff} F_8^{(9)}\right) \\  
\nn
&& C_{7,new}^{eff}= C_{7}^{(0) eff} + \as C_{7}^{(1) eff} + 
\as \left(4 C_{7}^{(0) eff} \omega_{7}(\hs) - C_{1}^{(0)} F_1^{(7)} -
C_{2}^{(0)} F_2^{(7)} - C_{8}^{(0) eff} F_8^{(7)} \right) \\ 
\nn
&& C_{10,new}^{eff}= C_{10}^{(0)} + \as C_{10}^{(1)} + 
\frac{\alpha_s}{\pi} C_{10}^{(0)} \omega_9(\hs) \\ 
\nn
&& C_{9,new}^{eff*}C_{7,new}^{eff} = C_{9}^{(0) eff^*}C_{7}^{(0) eff}
+\as \left(C_{9}^{(0) eff*}C_{7}^{(1) eff} + 
C_{9}^{(1) eff*}C_{7}^{(0) eff}\right) \\ 
\nn
&& + \as C_{9}^{(0) eff*}
\left(4 C_{7}^{(0) eff} \omega_{79}(\hs) - C_{1}^{(0)} F_1^{(7)} -
C_{2}^{(0)} F_2^{(7)} - C_{8}^{(0) eff} F_8^{(7)} \right) \\
&& + \as C_{7}^{(0) eff}
\left(4 C_{9}^{(0) eff*} \omega_{79}(\hs) - C_{1}^{(0)} F_1^{(9)*} -
C_{2}^{(0)} F_2^{(9)*} - C_{8}^{(0) eff} F_8^{(9*)}\right)
\eey
where the function $F_i^{j}$, $\omega_7$, $\omega_9$, $\omega_{79}$ are given in
\cite{13,7} and the relevant dimension six operators $O_i$ are the following:
\begin{eqnarray}
\nonumber
O_1 &=& \bar{s}_L \gamma^\mu T^a c_L \ \bar{c}_L \gamma_\mu T^a b_L,
\hspace{1.3cm}
O_{1u} = \bar{s}_L \gamma^\mu T^a u_L \ \bar{u}_L \gamma_\mu T^a b_L, \\
O_2 &=& \bar{s}_L \gamma^\mu c_L \ \bar{c}_L \gamma_\mu b_L,
\hspace{2.25cm}
O_{2u} = \bar{s}_L \gamma^\mu u_L \ \bar{u}_L \gamma_\mu b_L,  \\
\nonumber
O_7 &=& \frac{e}{16\pi^2} m_b(\mu) \ \bar{s}_L
\sigma^{\mu \nu} b_R \ F_{\mu \nu}, \hspace{0.55cm}
O_8=\frac{g_s}{16\pi^2} m_b(\mu) \ \bar{s}_L
\sigma^{\mu \nu} T^a b_R \ G_{\mu \nu}^a. \\
\nn
\nonumber
O_9 &=& \frac{e^2}{16 \pi^2} 
\bar{s}_L \gamma^{\mu} b_L \bar{e} \gamma_{\mu} e,
\hspace{2cm}
O_{10}=\frac{e^2}{16 \pi^2} \bar{s}_L \gamma^{\mu} b_L \bar{e} \gamma_{\mu} e
\end{eqnarray} 
In the expression for $C_{9,new}^{eff*}C_{7,new}^{eff}$
the \oass terms have been discarded. Similarly, in (4) for 
$|C_{i,new}^{eff}|^2$, i=7,9,10, only the terms linear in $\alpha_s$ 
are retained, when performing numerical calculations. 

\oas terms in (6) arise due to the \oas corrections to the Wilson coefficients
(the terms proportional to $C_i^{(1) eff}$ \cite{12}), 
due to the \oas corrections to 
$<e^+e^-s|O_i|b>$ matrix elements and due to the singular graphs connected with
the gluon bremsstrahlung ($b \to s e^+ e^- g$) effects \cite{13}. The
contribution of nonsingular bremsstrahlung diagrams and \oas corrections to the
$O_{1u}$ $O_{2u}$ matrix elements are still unknown. It is expected however for
them to be small.

Both for the SM and the 2HDM the content of the operators $O_i$ is the same, so
is the anomalous dimension matrix. The new physics effects enter only through
the matching conditions. In other words, formulae (4) and (5) are valid also in
the 2HDM, however for the Wilson coefficients one has now \cite{14,18,19,20}
\bey
C_{1,2}^{(0)}(\mu)&=&C_{1,2}^{(0)^{SM}}(\mu) \\ 
C_{7,8}^{eff}(\mu)&=&C_{7,8}^{eff^{SM}}(\mu) - \xi_t \xi_b
C_{7,8}^{eff^{H}}(\mu) + |\xi_t|^2 C_{7,8}^{eff^{\tilde{H}}}(\mu) \\ 
C_{9,10}^{eff}(\mu)&=&C_{9,10}^{eff^{SM}}(\mu) + 
|\xi_t|^2 C_{9,10}^{eff^{\tilde{H}}}(\mu)  
\eey
The full set of formulae for $C_{7,8}^{eff}(\mu)$ (so is for 
$C_{1,2}^{(0)}(\mu)$ in the chosen operator basis) is given in \cite{18}. Here 
it is worth to notify only that the contribution of the second term in r.h.s. 
of (8) dominates over the last term\footnote{Moreover, the last term in (8) 
is about one order smaller than the SM contribution.}. For 
$C_{9,10}^{eff^{SM}}(\mu)$ one has
\bey
\nn
C_{9}^{eff^{SM}}(\mu)&=&\tilde{C}_{9}^{eff^t}(\hs) - \tilde{C}_{9}^{eff^c}(\hs)
+ \frac{\lambda_u^s}{\lambda_t^s} \Delta \tilde{C}_{9}^{eff}(\hs) \\
C_{10}^{eff^{SM}}(\mu)&=&\tilde{C}_{10}^{eff^t}(\hs) - 
\tilde{C}_{10}^{eff^c}(\hs)  
\eey
where $\tilde{C}_{9,10}^{eff^t}(\hs)$, $\tilde{C}_{9,10}^{eff^c}(\hs)$ and
$\Delta \tilde{C}_{9}^{eff}(\hs)$ are given in \cite{12}.
For $C_{9,10}^{(0)eff^{\tilde{H}}}(\mu)$ one can easy deduce that
\bey
\nn
C_9^{(0)eff^{\tilde{H}}}(\mu) &=& C_9^{(0)^{\tilde{H}}}(\mu_W) +
C_4^{(1)^{\tilde{H}}}(\mu_W) \sum_{i=5}^{9}{q_i^{t(+)} 
\tilde{\eta}^{a_i + 1}} \\
C_{10}^{(0)eff^{\tilde{H}}}(\mu) &=& C_{10}^{(0)^{\tilde{H}}}(\mu_W)  
\eey
where $\tilde{\eta}=\alpha_s(\mu_W)/\alpha_s(\mu)$, 
$C_{9,10}^{(0)^{\tilde{H}}}(\mu_W)$ are given in \cite{19,20}, 
$C_4^{(1)^{\tilde{H}}}(\mu_W)$ is given in \cite{18} and the "magic numbers"
$q_i^{t(+)}$ can  be found in \cite{12}.

Unfortunately the matching conditions for $C_{9,10}^{(1)eff^{\tilde{H}}}$ are
unknown yet. For this reason here the SM values 
$C_{9,10}^{(1)eff^{\tilde{H}}}(\mu)$ will be used. \\
\\
5. The effective Wilson coefficients\footnote{Because of the smallness of 
$C_8^{(0)eff}$ and condition (7) the qualitative discussion throughout the
paper is valid both for the old and
the new Wilson coefficients, unless the difference is specially notified.}
deviate from their SM values differently. Thus $C_9^{eff}$ deviates for
$C_{9}^{eff^{SM}}$ only by few percents. On the contrary, $|C_{10}^{eff}|$ can 
be about 1.4 times larger than in the SM\footnote{One may subsequently expect 
that
omitted here \oas corrections to $C_{10}^{eff}$ will not exceed 
$(10 \div 15)\%$ and those to $C_9^{eff}$ will be negligible.} (Table~1). 

\begin{table}
\caption{$R_{10}=C_{10}^{(0)^{\tilde{H}}}/C_{10}^{(0)^{SM}}$.}
\begin{center}
\begin{tabular}{|c|c|c|c|}
\hline
$m_{H^+}$ & 100GeV & 200GeV & 400GeV \\
\hline 
$R_{10}$ & $0.36 \div 0.39$ & $0.23 \div 0.26$ & $0.12 \div 0.14$ \\
\hline  
\end{tabular}
\end{center}
\end{table}
The largest deviations from the SM interval occur for $C_7^{eff}$. It is known
that in the standard model extensions $Re(C_7^{eff})$ may have the sign 
opposite 
to that in the SM and furthermore unlike the SM the dispersive part of 
$C_7^{eff}$ can be 
complex (recall that in the SM $C_{7}^{(0)eff}$ is real and 
the imaginary part of $C_{7,new}^{eff}$ arises 
only due to absorptive parts of the $O_1$ and $O_2$ matrix elements). As for
$|C_7^{eff}|$, it is bounded due to the experimental constraints on the $\bxg$
branching. In this paper the numerical calculations are performed both 
including and
neglecting \oas effects, and the derived results are compared to each 
other.
When neglecting \oas corrections to $B(\bxee)$ is it reasonable to do the same
also for $B(\bxg)$. In this case more conservative bound than (1) should be used:
one takes therefore \cite{18}
$1 \times 10^{-4} < B(\bxg) < 4.2 \times 10^{-4}$. This gives
$0.04 \leq |C_7^{(0) eff}|^2 \leq 0.18$ in the next-to-leading order. 
When including \oas
corrections, the condition (1) puts some constraint on $|C_{7,new}^{eff}|^2$ 
at the
point $\hs=0$. Though this point is out of the consideration, due to the weak
dependence of $C_{7,new}^{eff}$ on $\hs$ (see formulae of \cite{13}) the
aforesaid constraint is essential also for the considered range of the dilepton
invariant mass.

Because of discarding \oass terms in the expression for $|C_{7,new}^{eff}|^2$,
there are some problems, connected with the bounds on this quantity.
\begin{table}[t]
\caption{The restrictions on $|C_{7,new}^{eff}(\mu)|^2_{|\hs=0}$ 
for $m_{H^+}=100$GeV, coming from the
condition $B(B \to X_s \gamma)=(2.96 \pm 0.35) \times 10^{-4}$.}
\begin{center}
\begin{tabular}{|c|c|c|c|c|c|c|}
\hline
 & \multicolumn{3}{|c|}{$\mu_W=M_W$} & \multicolumn{3}{|c|}{$\mu_W=m_t$} \\
\hline
 & $\mu=m_b/2$ & $\mu=m_b/2$ & $\mu = 2 m_b$
& $\mu=m_b/2$ & $\mu=m_b/2$ & $\mu = 2 m_b$ \\
\hline
$|C_{7,new}^{eff}|^2$ & -.04 $\div$ 0.13 & .03 $\div$ .12 & .05 $\div$ .12
& .05 $\div$ .13 & .05 $\div$ .12 & .06 $\div$ .12 \\
\hline
\end{tabular}
\end{center}
\end{table}
As on can see from the Table~2, for $\mu_W=m_W$ the restrictions on 
$|C_{7,new}^{eff}|^2_{\hs=0}$ are highly sensitive to the choice of the 
low-energy scale. Moreover, for $\mu=m_b/2$
the condition (1) allows negative values of
$|C_{7,new}^{eff}|^2$. 
The unnatural negative values of $|C_{7,new}^{eff}|^2$ indicate
on the fact that in some regions of the 2HDM parameter space the discarded 
\oass terms are numerically relevant and may not be neglected. 
Such a situation occurs in particular in the case when $|C_7^{(0)eff}| \sim 1$,
however due to the \oas corrections  $|C_{7,new}^{eff}| \sim
|C_{7,new}^{eff^{SM}}|$ or even smaller. Then it is possible that although
$\Gamma(b \to s\gamma) \sim  |C_{7,new}^{eff}|^2 < 0$, the long-distance 
$O(1/m_b^2)$ corrections
(which to the considered order are taken proportional to $|C_7^{(0)eff}|^2$
\cite{18})
drive $\bxg$ branching to the experimentally allowed interval. 

Of course there is no sense to consider such unphysical possibilities, 
which arise only due to neglecting of higher order corrections to the $\bxg$
branching. One can easy avoid to consider such unfavorable regions of 
the parameter space.
As one can see from the Table~2, for $\mu_W=m_t$ the reliability of 
the restrictions on $|C_{7,new}^{eff}|^2$ increases. In particular, the allowed
interval of $|C_{7,new}^{eff}|^2_{\hs=0}$ is weakly sensitive to the choice
of the low-energy scale and one may take now
$0.05 \leq |C_{7,new}^{eff}|^2_{\hs=0} \leq 0.13$ (this bound turns to be valid
also for $m_{H^+}=200$GeV and $m_{H^+}=400$GeV). This result is not surprising:
it is known that for $\mu_W=m_t$, \oas
corrections to the $\bxg$ branching are minimized, as compared to the case of
$\mu_W=M_W$ \cite{18}. Thus the above-mentioned problem is avoided, if 
calculations are performed only for
$\mu_W=m_t$. This strategy will be used throughout this paper in the most of
cases. However, the
estimation of possible inaccuracy in the obtained results, connected with the
uncertainty in the choice of the matching scale, will be done as well. \\
\\
6. Let us proceed to the numerical results. During the numerical analysis 
we use  $M_W=80$GeV,
$m_t^{pole}=(174 \pm 5)$GeV, $\alpha_s(M_Z)=0.119 \pm 0.002$ \cite{21},
$m_b^{pole}=(4.8 \pm 0.15)$GeV, $m_c/m_b=0.29 \pm 0.02$, 
$\alpha_{em}=1/(130.3 \pm 2.3)$ and $B_{sl}=
(10.49 \pm 0.46)\%$ (see \cite{16} and references
therein). 
The low-energy scale $\mu$ is varied as $\mu=m_b/2$, $\mu=m_b$, 
$\mu=2m_b$. As it was noted already, the matching scale is identified here
with the top quark mass. Only when the heavy mass scale dependence is examined,
the matching scale is chosen  as $\mu_W=M_W$ as well.

The charged Higgs mass is varied as $m_{H^+}=100$GeV, 200GeV, 400GeV.
The restrictions on $|\xi_t|$ and $|\xi_b|$ are derived from the
requirement for the top and bottom Yukawa couplings to be in the perturbativity
range in a whole energy interval between the electroweak breaking scale and
unification scales, and using the experimental constraints coming from the
measurements of $\bxg$ branching ratio and $B - \bar{B}$ mixing effects (see
the discussion in previous section and
\cite{16} for more details).

At the low-energy scale $\alpha_s$ is computed, using its two-loop 
renormalization group equation \cite{18}. However, when neglecting \oas 
corrections to $\bxee$ and $\bxg$ branching ratios, one-loop result for 
$\alpha_s(\mu)$ is used.

In the Wolfenstein parameterization  the necessary CKM-factors are given by
\cite{22}:
\beq
\frac{\lambda_t^s}{|V_{cb}|} = -(1- \lambda^2/2 + \lambda^2 (\rho -
i \eta)),
\hspace{0.5cm}\frac{\lambda_u^s}{|V_{ub}|} = \lambda^2 (\rho - i \eta),
\eeq
where $\lambda=\sin{\theta_C} \approx 0.22$ and the unitarity triangle 
parameters $\rho$ and $\eta$ can be obtained from
the unitarity fits, which yield \cite{23,24}
\beq
\sqrt{\rho^2 + \eta^2} = 0.423 \pm 0.064, \hspace{0.5cm}
38^o \leq {\bar\gamma}=\arctan{\frac{\eta}{\rho}} \leq 81^o  
\eeq
The numerical calculations consist of two steps. At first one investigates the 
renormalization scale dependence of the $\bxee$ normalized differential 
width: 
\begin{displaymath}
R(\hs)=\frac{1}{\Gamma(B \to X_c e^+\nu)}\frac{d\Gamma(\bxee)}{d\hs}
=\frac{1}{B_{sl}}\frac{dB(\bxee)}{d\hs}
\end{displaymath} 
The use of $R(\hs)$ would allow one to compare our results with those of 
ref.'s \cite{12,13}. The renormalization scale dependence is examined for 
some fixed characteristic values of the new physics parameters, for the 
"best fit" values of the CKM parameters ($\rho=0.19$, $\eta=0.37$) and for 
the central values of the remaining parameters of theory.

The next step involves complete investigation for the differential and 
partially integrated branching ratios of $\bxee$ in the 2HDM and comparison 
of the obtained results with those in the SM. The parameters of theory 
are varied now in the intervals specified above.

Let me briefly recall how the situation with the low-energy and 
matching scale dependence of $R(\hs)$ in the SM looks. On the absence of the 
\oas 
contribution (then $C_{i,new}^{eff}$ in (4) are replaced by 
$C_i^{(0)eff}$, i=7,9,10), 
the $\mu$-dependence of $R(\hs)$ (in the naive scheme) is $\sim 6\%$ or 
smaller. 
However, such a weak sensitivity of the $\bxee$ decay width to the choice of 
the low-energy scale in the next-to-leading 
order is accidental \cite{8}. The error from the $\mu$-dependence of 
$R(\hs)$ grows up to 13\%, when \oas corrections to the Wilson 
coefficients are taken into account \cite{12}. Only after \oas 
corrections to the
matrix elements of the operators $O_j$, j=1,2,7,..,10, are also included, this 
error is reduced down to 6.5\% \cite{13}. As for the error, connected with the
uncertainty of the heavy mass scale, it is reduced from $\sim 15\%$ in the NLO
to few percents in the NNLO\footnote{Although this result has been derived in
\cite{12} considering \oas corrections only to the Wilson coefficients, it is
easy to check that it remains valid also after taking into account \oas
corrections to the operators matrix elements.}. 
\begin{table}
\caption{The matching scale dependence of
$C_{10}(\mu)=C_{10}^{(0)}(\mu) + \alpha_s(\mu)/(4 \pi)\ C_{10}^{(1)}(\mu)$ 
in the SM and in the 2HDM for
$|\xi_t|=1$ and $m_{H^+}=100$GeV.}
\begin{center}
\begin{tabular}{|c|c|c|c|c|}
\hline
 & \multicolumn{2}{|c|}{with \oas corrections} 
& \multicolumn{2}{|c|}{without \oas corrections} \\
\hline
 & $\mu_W=M_W$ &$\mu_W=m_t$ & $\mu_W=M_W$ &$\mu_W=m_t$\\
\hline
$C_{10}^{eff^{SM}}(\mu)$ & -4,37 $\div$ -4.02 & -4.50 $\div$  -4.10 &
-4.78 $\div$ -4.37 & -4.46 $\div$ -4.05 \\
\hline
$C_{10}^{eff}(\mu)$ & -6.24 $\div$ -5.64 & -6.17 $\div$ -5.54  & 
-6.65 $\div$ -6.00 & -6.13 $\div$ -5.50 \\
\hline
\end{tabular}
\end{center}
\end{table}

The difference between the renormalization scale behavior of $R(\hs)$ in 
the SM and 2HDM occurs predominantly due to $C_{7}^{eff}$. The $\mu$- 
dependence of $C_{10}^{eff}$ originates only at \oas order and is 
small therefore. There is no essential 
difference between the $\mu_W$ - dependence of $C_{10}$ in the SM 
and 2HDM even for $|\xi_t| \sim 1$. As one can see from the Table~III, in both
of models the $\mu_W$ - error of $C_{10}$ is $\sim 10\%$ in the NLO 
and $\sim (2-3)\%$ in the NNLO.

When investigating the renormalization scale behavior of $R(\hs)$, the new 
physics parameters are chosen as $m_{H^+}=100$GeV, $|\xi_t|^2=0.5$ and 
a) $\xi_t \xi_b=6.4$, b) $\xi_t \xi_b = 2.8 \pm 2.8 i$. In the case a), 
the dispersive part of $C_{7}^{eff}$ is real and has the sign opposite to that 
in the 
standard model. In the case b), the imaginary part of $C_{7}^{eff}$ 
dominates over the real one\footnote{For the cases a) and b) the parameters of 
theory are within or close to the experimentally allowed values only in 
the NNLO. As it is shown in the section~8, the experimentally allowed 
intervals of the new physics parameters are strongly different in the 
NLO and the NNLO.}.
\begin{figure}[t] 
\vspace{-1cm}
\begin{flushleft}
\leavevmode
\epsfysize=12.5cm
\mbox{\hskip -0.25in \epsfbox{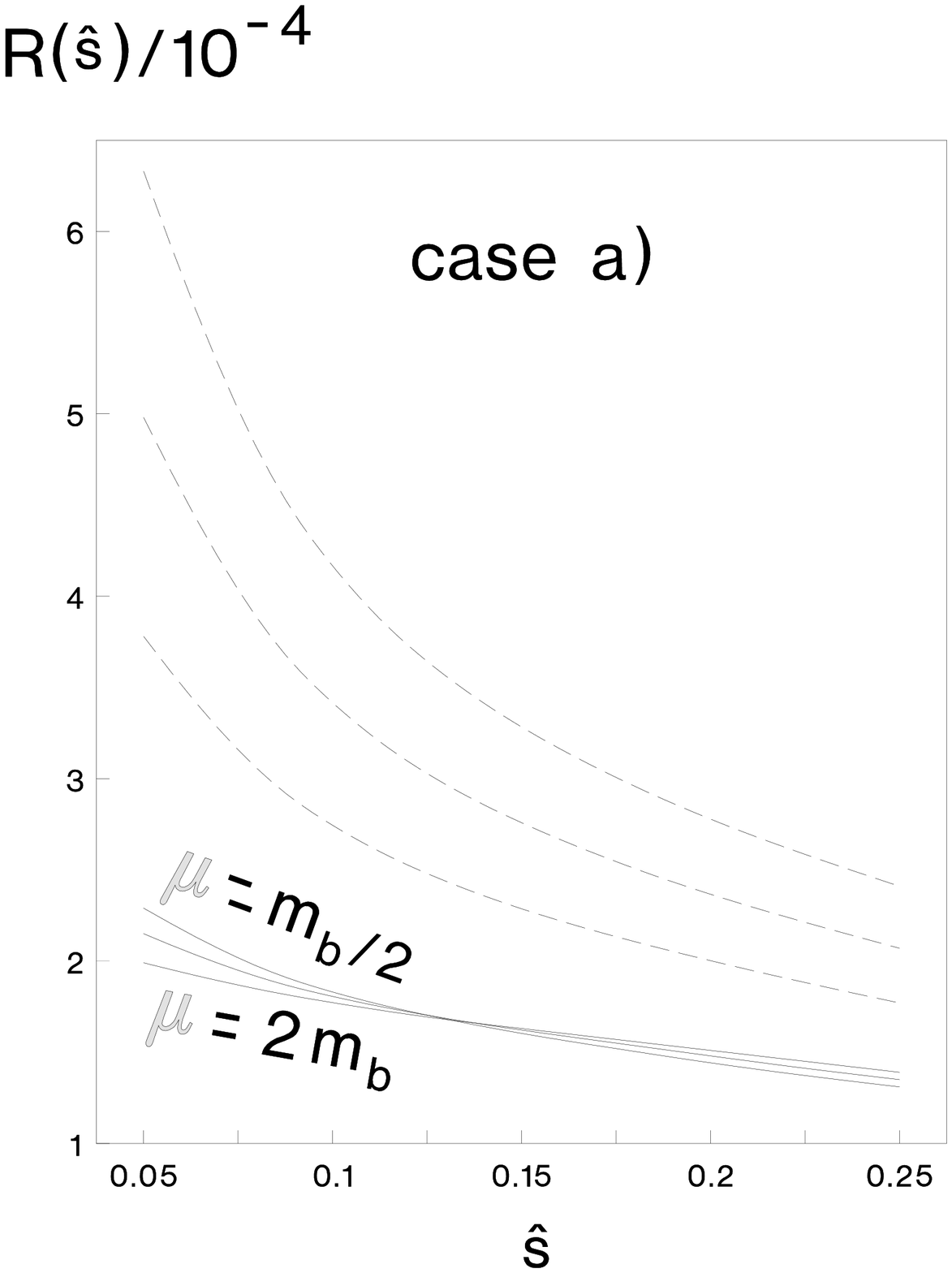}}
\end{flushleft}
\vspace{-13.45cm}
\begin{flushright}
\leavevmode
\epsfysize=12.5cm
\mbox{\hskip -0.25 in \epsfbox{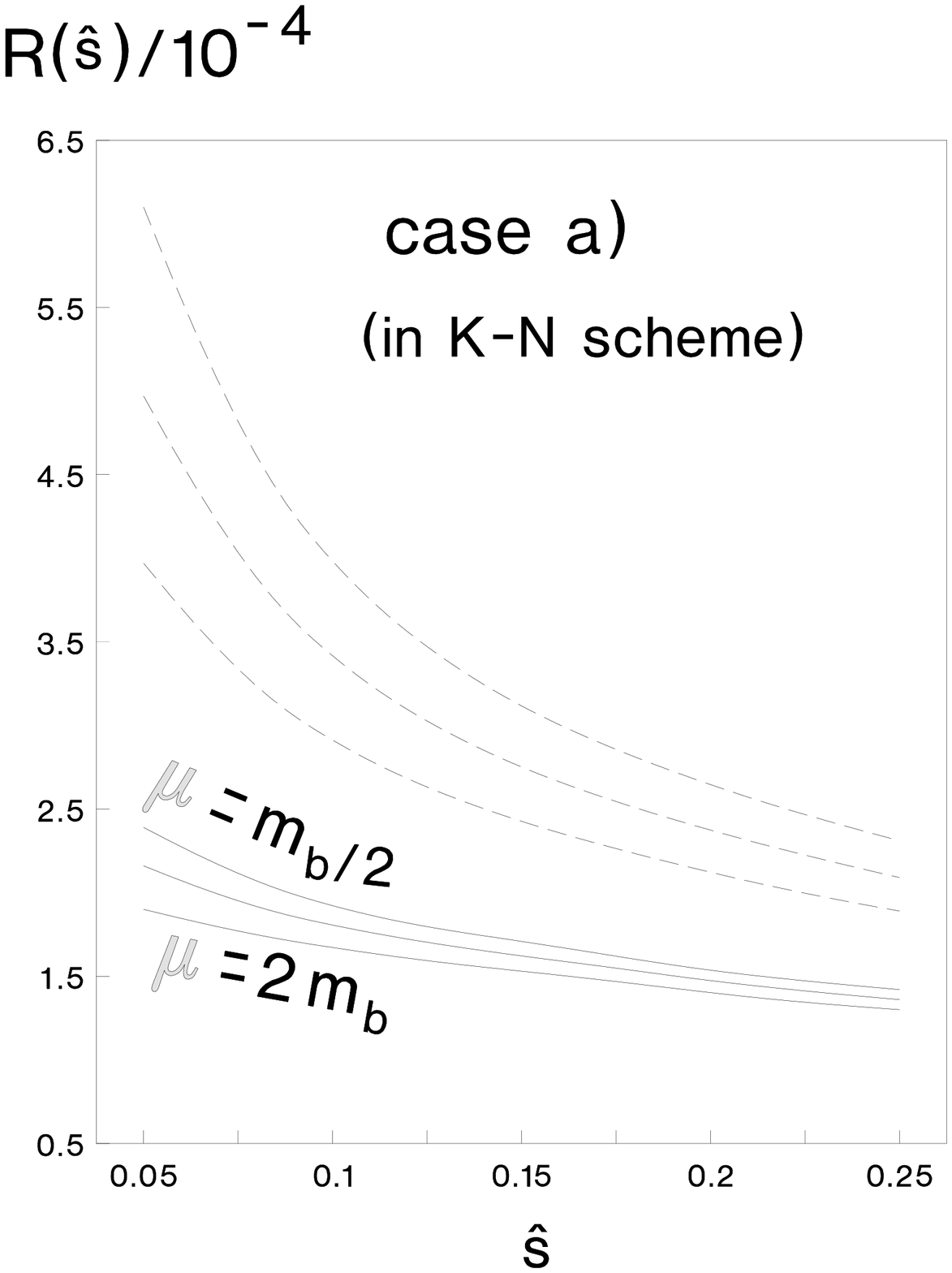}}
\end{flushright}
\ve{-1.cm}
\caption{Low energy scale dependence of 
$R(\hs)$ in the NNLO (solid lines) and NLO (dashed and dotted lines) 
in the 2HDM, case a), for $\mu_W=m_t$. The analysis is performed both in the
naive and the Kagan-Neubert schemes.
Unless notified, lower dashed (dotted)
and solid lines correspond to $\mu=m_b/2$, middle lines correspond to
$\mu=m_b$ and upper lines correspond to $\mu=2 m_b$.} 
\end{figure}

For the case a), the behavior of the $\bxee$ normalized differential 
width as a function of the dilepton invariant mass for different choices of the
low-energy scale and for $\mu_W=m_t$ is presented in the Fig.~1. As one can see
from this figure, in the next-to-leading order the uncertainty of $R(\hs)$
connected with that of the low-energy scale is large enough: it reaches 25\%
for lower values of $\hs$ and 17\% for larger values of $\hs$. \oas corrections
reduce the $\mu$-error of $R(\hs)$ significantly: in the naive scheme (where the
errors of different terms in (4) are simply summed) it is $\sim 10\%$ for lower
values of $\hs$ and its sign is flipped, it is only about few percents for
larger values of $\hs$, and it almost disappears for the intermediate values
of $\hs$. Such a dependence of $R(\hs)$ on the choice of the low-energy scale
indicates on the cancellation of $\mu$-errors of the terms in (4). In such cases
one usually uses so-called Kagan-Neubert method \cite{25,8}: 
in context of the present calculations this approach implies the calculation of
$\mu$-errors of different blocks of the (effective) Wilson coefficients 
separately and
then adding them in quadrature. The result derived in the Kagan-Neubert scheme 
does not differ essentially from that in the naive scheme: now the uncertainty 
of $R(\hs)$, connected with that of the low-energy scale, is $\sim 12\%$ for
lower values of $\hs$ and $\sim 6\%$ for large values of $\hs$. This means that
weak $\mu$-dependence of $R(\hs)$ in the NNLO is not accidental.

The investigations in the Kagan-Neubert scheme show also that in the
next-to-next-to-leading order the main source of the $\mu$-dependence is the
term proportional to 
$4\left(1 + \frac{2}{s}\right) |C_{7,new}^{eff}(\mu)|^2$
(in the NLO there is also sizable contribution from the term proportional to
$Re\left(C_{9,new}^{eff*}(\mu) C_{7,new}^{eff}(\mu) \right))$. This explains 
why in the NNLO the low-energy scale dependence of $R(\hs)$ 
for larger values of $\hs$ is especially small.
\begin{figure}[t]
\vspace{-1cm}
\begin{center}
\leavevmode
\epsfysize=12.5cm
\mbox{\hskip -0.25 in \epsfbox{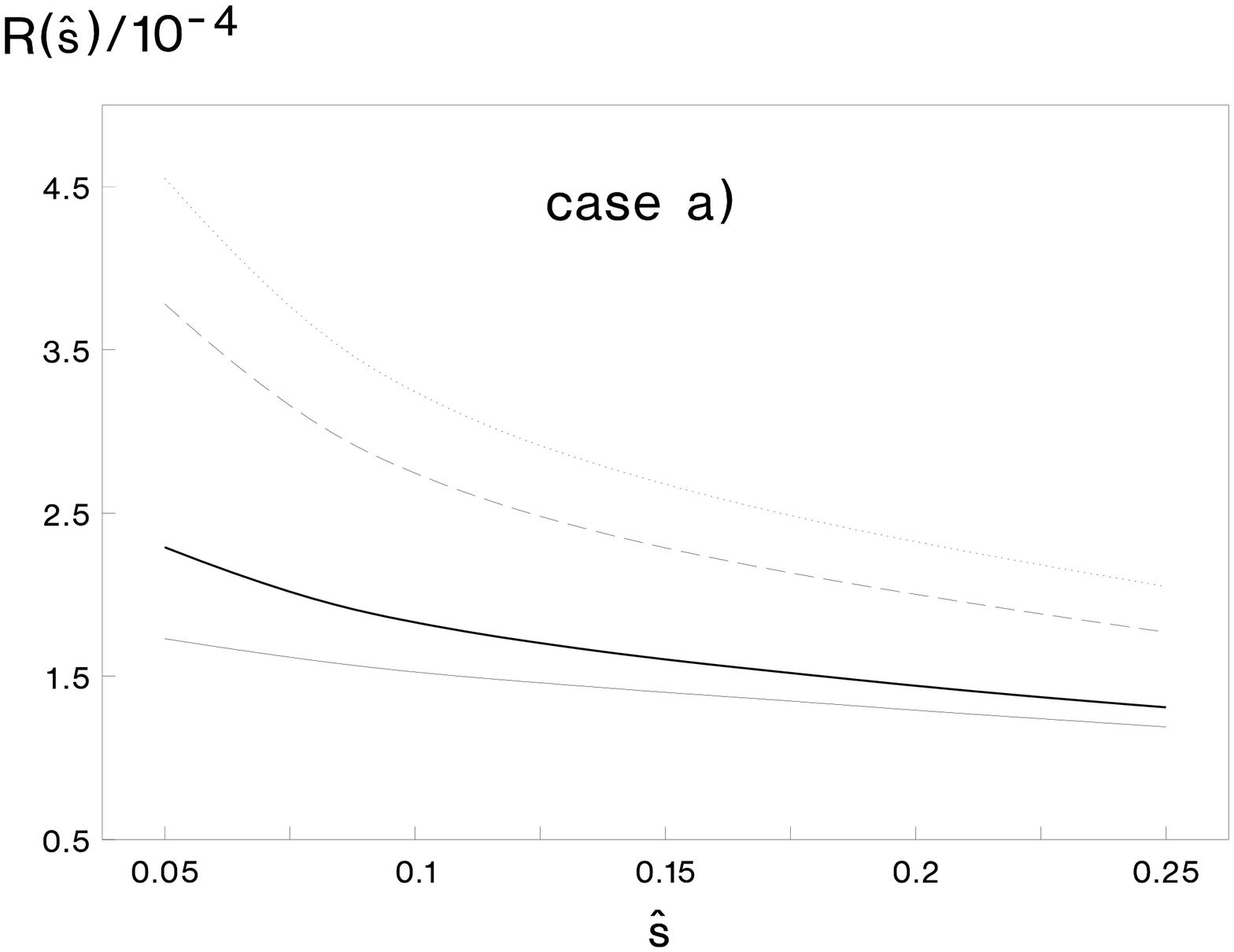}}
\end{center}
\vspace{-0.5cm}
\caption{The matching scale dependence of
$R(\hs)$ in the NNLO (solid lines) and NLO (dashed and dotted lines) 
in the case a) for
$\mu=m_b/2$. The fat (thin) solid and dashed (dotted) lines correspond to 
$\mu_W=m_t$ ($\mu_W=M_W$).}
\end{figure}
  
It is important to stress that in the case a), \oas corrections suppress the
$\bxee$ decay width about $1.5 \div 3$ times, depending on $\hs$ and $\mu$. 
In other
words, NNLO corrections to $R(\hs)$ are about $(35 \div 65)\%$ of the leading
and next-to-leading order terms. It is reasonable therefore to expect that 
\oass
contribution to the $\bxee$ decay width will be nonnegligible as well. This
is argued also by the investigation of the matching scale dependence of 
$R(\hs)$. As one can see from the Fig.~2, in the next-to-leading order the
uncertainty of $R(\hs)$ is $\sim 15\%$, when varying the matching scale from
$m_t$ to $M_W$. This uncertainty is almost independent on
$\hs$. In the next-to-next-to-leading order the situation is quite different:
while for larger values of $\hs$, $\mu_W$-error of $R(\hs)$ decreases down to
$\sim 10\%$, for lower values of $\hs$ it increases up to $\sim 25\%$. Such a
large error can be reduced only by higher order corrections to the $\bxee$ decay
width. 
\begin{figure}[t]
\vspace{-1cm}
\begin{flushleft}
\leavevmode
\epsfysize=12.5cm
\mbox{\hskip -0.25 in \epsfbox{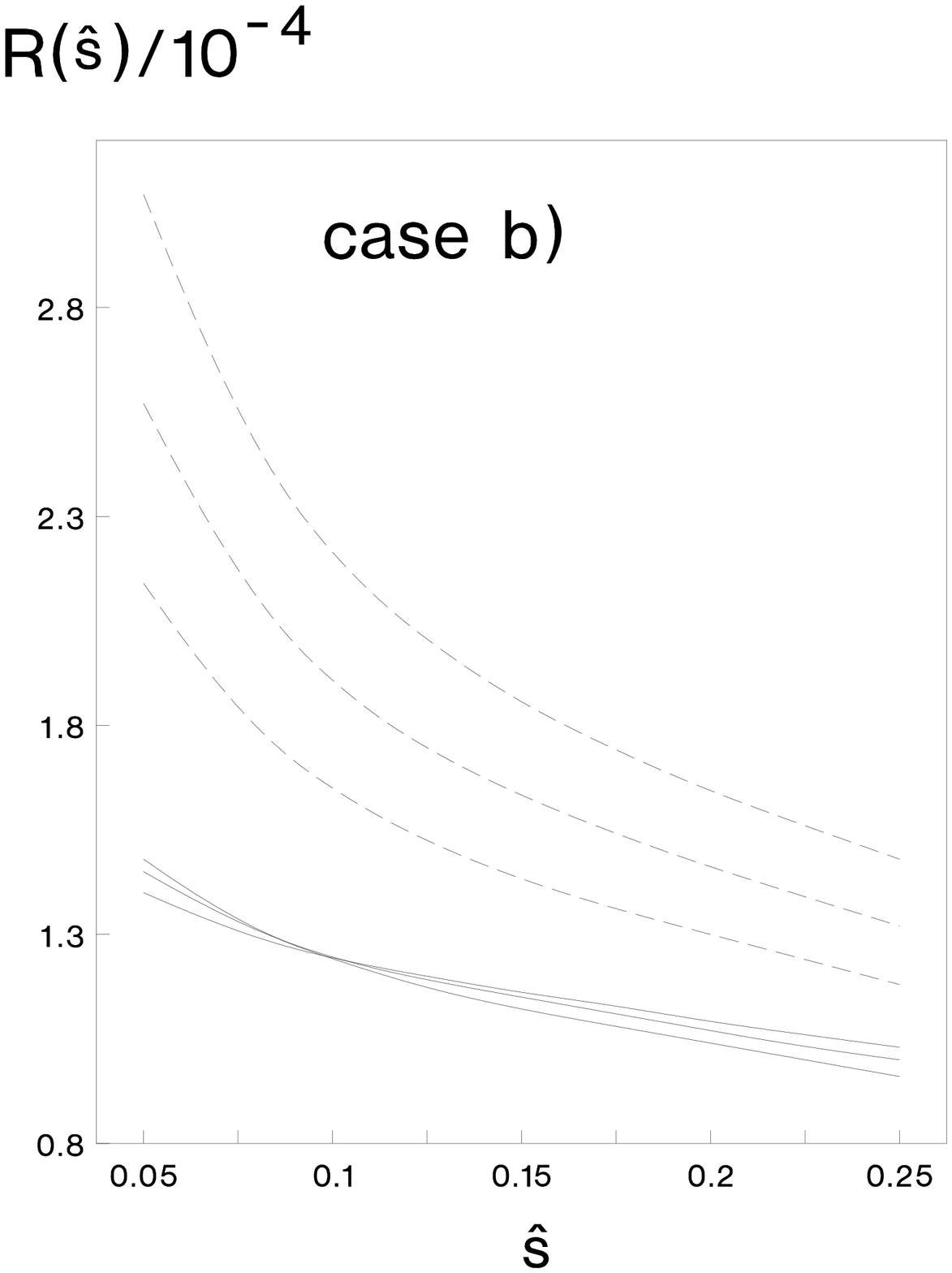}}
\end{flushleft}
\vspace{-13.45cm}
\begin{flushright}
\leavevmode
\epsfysize=12.5cm
\mbox{\hskip -0.25 in \epsfbox{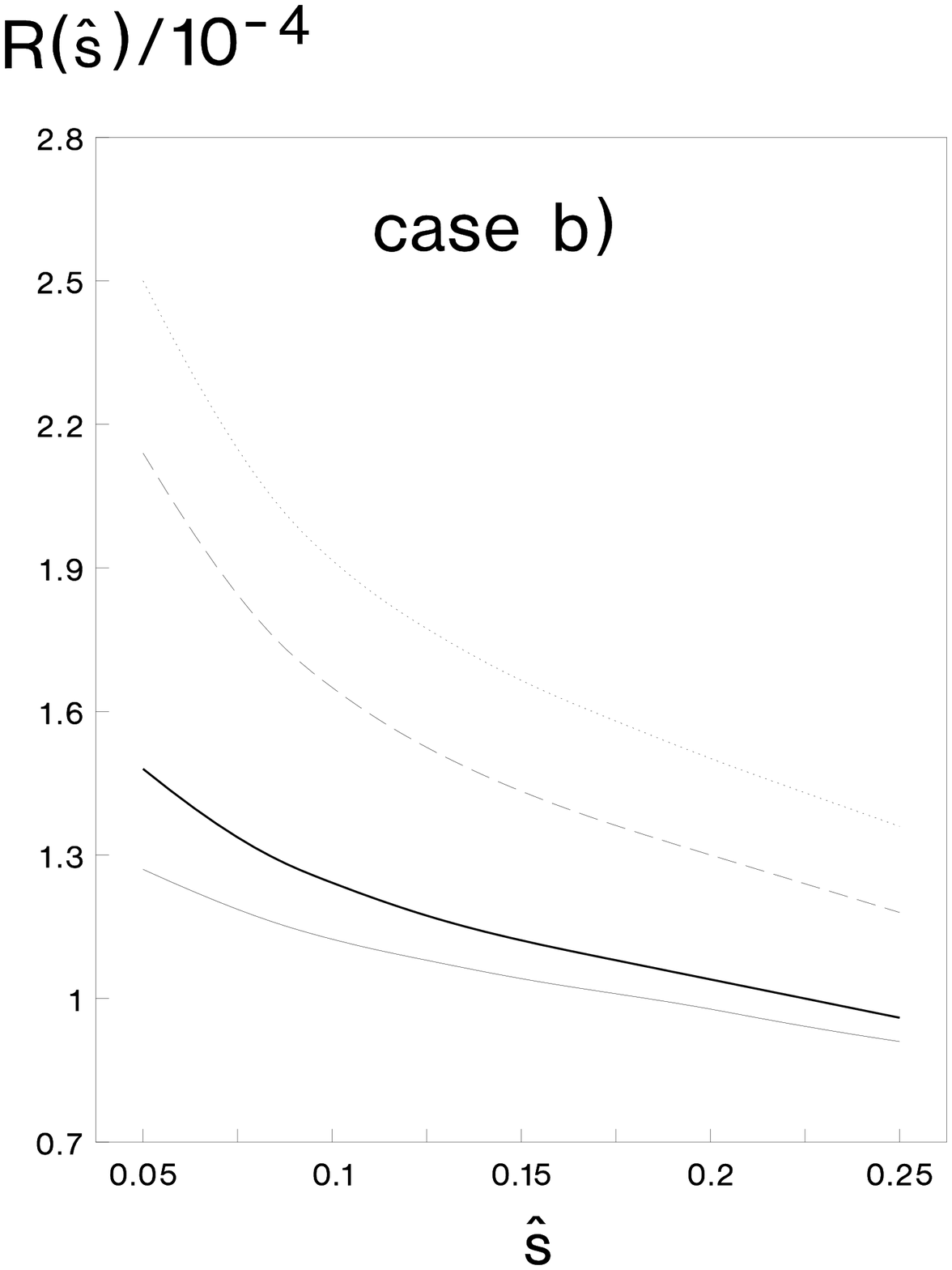}}
\end{flushright}
\ve{-1cm}
\caption{The same as in Fig.'s 1,2 (respectively the left- and 
right-hand-side graphs) but for the case b). } \end{figure}

The obtained results for the case~b) are presented in the Fig.~3. As one 
can see from this figure, the situation is similar to that of the 
case~a). Due to the \oas corrections, the low-energy scale 
dependence of $R(\hs)$ becomes negligible, so is the matching scale 
dependence for larger values of $\hs$. However for lower values of $\hs$, 
the $\mu_W$-error of $R(\hs)$ is of the same order ($\sim 14\%$) as 
without \oas corrections.

Thus, although
in the 2HDM \oas corrections reduce significantly the 
low-energy scale
dependence of the $\bxee$ decay width, higher order terms in the perturbation 
theory can still be nonnegligible, when  
the new physics contribution is sizable. The importance of higher order 
corrections is manifested by the sensitivity of the obtained results to 
the choice of the heavy mass scale. \\
\\
7. Let us compare now the predictions of the 2HDM for the $\bxee$ branching 
ratio 
to those of the SM (whole allowed range of the 2HDM parameter space is 
considered 
now). The deviation of $B(\bxee)$ from the SM results can take
place due to the change of the sign of $Re(C_{7}^{eff})$ (the source~I), 
due to the
imaginary part of $C_{7}^{eff}$, connected with the Higgs doublet vacuum phase
(the source~II), and due to the deviation of $C_{10}^{eff}$ from
$C_{10}^{eff^{SM}}$ (the source~III). It is easy deduce from the formula (4)
that all aforementioned three sources increase the $\bxee$ branching, once
$Re(C_{7}^{eff^{SM}}) < 0$ and $|C_{10}^{eff}| \geq |C_{10}^{eff^{SM}}|$.
 
The contribution of the source~II is expected to be small: it is proportional 
to\footnote{During the numerical calculation the terms which make the difference
between the $\bxee$ decay and the CP-conjugated decay are dropped.}
$Im(C_{9,new}^{eff})Im(C_{7,new}^{eff})$ and one has 
$|Im(C_{9,new}^{eff})| \ll |Re(C_{9,new}^{eff})|$. The effects connected with 
the
source~III are strongly correlated with the sign of $Re(C_{7}^{eff})$. In the
SM-like case the terms of (4), proportional to $|C_{9,new}^{eff}|^2$ and 
$Re(C_{9,new}^{eff*} C_{7,new}^{eff})$, have opposite signs so that their
contribution is partially canceled. Consequently, the partial weight of the
the term proportional to the $|C_{10,new}^{eff}|^2$ and hence the contribution
of the source~III is significant. Respectively, when $Re(C_{7}^{eff})$ has the
sign, opposite to that in the standard model, the effects of the source~III are
minimized. It is straightforward to deduce from current discussion that the
sources~I~and ~III do not interfere in fact.
\begin{figure}[t]
\ve{-1cm}
\begin{center}
\leavevmode
\epsfysize=12.5cm  
\mbox{\hskip -0.3 in \epsfbox{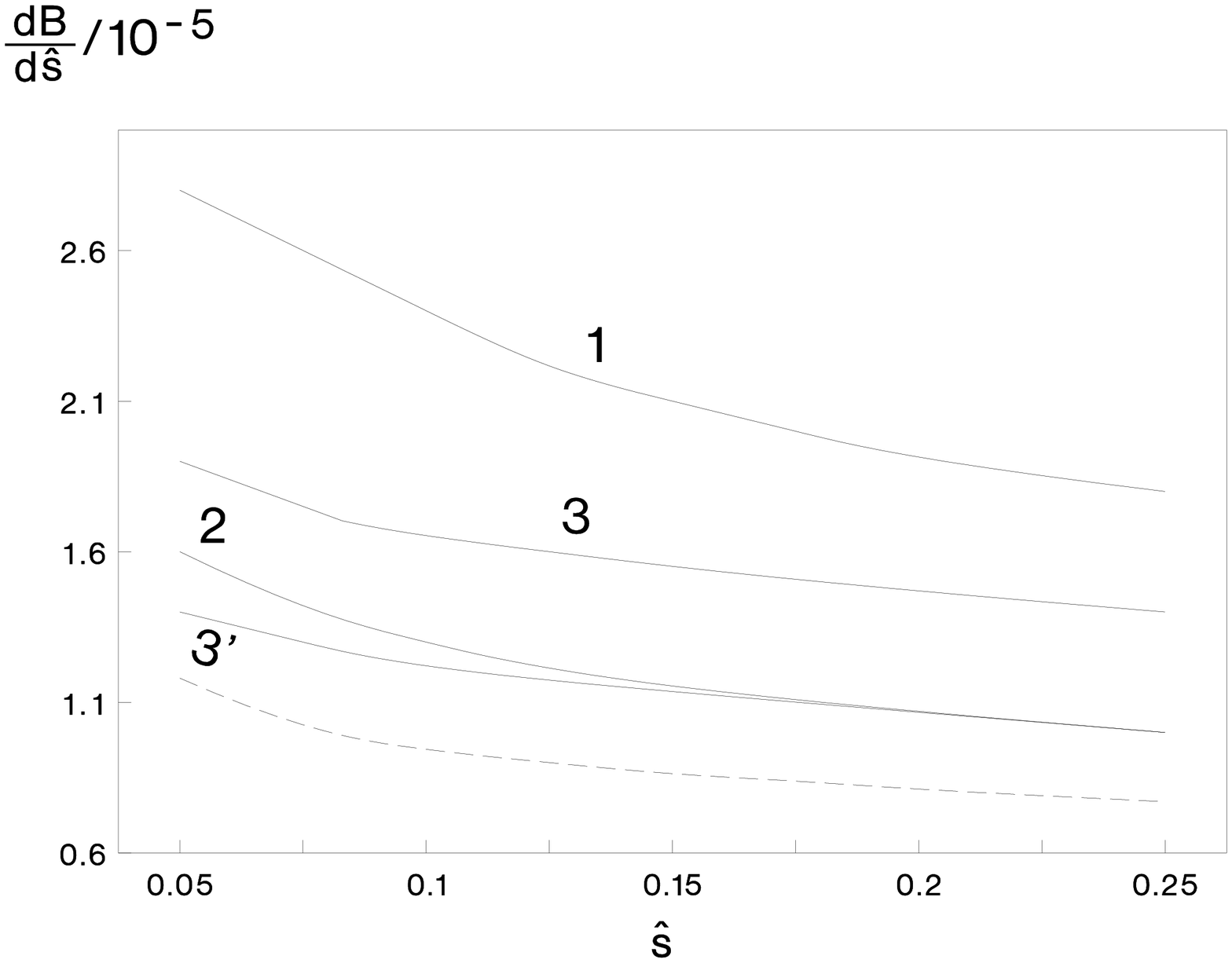}}
\end{center}
\ve{-0.5cm}
\caption{Maximum value of 
$dB(\bxee)/d\hs$ in the SM (dashed line) and 2HDM (solid lines). 
The following regions of the 2HDM parameter space are considered: 
$Im(\xi_t \xi_b)=0$, $|C_{10,new}^{eff}|^2 <
1.15 |C_{10,new}^{eff^{SM}}|^2$ (line 1, only the source~I is actual); 
$sign(Re(C_{7,new}^{eff}))=sign(Re(C_{7,new}^{SM^{eff}}))$, 
$|C_{10,new}^{eff}|^2 <
1.15 |C_{10,new}^{SM^{eff}}|^2$ (line 2, only the source~II is actual); 
$Im(\xi_t \xi_b)=0$,
$sign(Re(C_{7,new}^{eff}))=sign(Re(C_{7,new}^{SM^{eff}})$, 
$m_{H^+}=100$GeV, 400GeV
(lines 3, 3' respectively, only the source~III is actual).}
\end{figure}

The maximum value of the $\bxee$ differential branching ratio as a function 
of $\hs$
is presented in the Fig.~4 for particular cases, when only one of the
aforementioned sources is actual. As it was expected, the contribution of the
source~II is not large: if only this source is actual,  
$dB(\bxee)/d\hs$ deviates from its SM maximum value at most 1.4 times. 
The source~III can make $dB(\bxee)/d\hs$ $1.6 \div 1.9$
times larger than in the SM (respectively for $\hs$ varying from 0.05 to 0.25).
The deviations from the SM results,  connected with the source~III, are most
perceptible when $|\xi_b| \ll 1$ and $|\xi_t| \sim 1$. The effects of the 
source~III are rapidly
minimized with the increasing of the charged Higgs mass: for $m_{H^+}=400$GeV
they are already of the same order as those connected with the source~II.

The most sizable deviations from the SM predictions occur due to the source~I.
Due to this source, maximum value of $dB(\bxee)/d\hs$ can be up to 2.6 times
larger than in the standard model. The effects connected with the
source~I are almost insensitive to the charged Higgs mass. 
Large deviations from the SM result are possible even for $m_{H^+} \sim 1$TeV
(until $Re(C_7^{eff})$ becomes positive, when $|\xi_b| \gg 1$). 
However
the consideration of such large values of the charged Higgs mass is out of the
scope of the present paper, otherwise one should take into account the
difference between the electroweak breaking scale and the charged Higgs mass
scale.

It is worth to make here some digression on the situation in the
Models~I~and~II. If one applies the restrictions on the parameters $\xi_t$ and
$\xi_b$ notified in the section~6, only the source~III will be relevant
in these versions of the 2HDM. But the effects of this source are very small, 
because one has $|\xi_t|^2 < 1/4$ for $m_{H^+}=100$GeV in the Model~I and
$m_{H^+} > 400$GeV in the Model~II. As a result, the predictions of these models
for $\bxee$ branching are close to those of the SM.
\begin{figure}[t]
\ve{-1cm}
\begin{flushleft}
\leavevmode
\epsfysize=12.5cm
\mbox{\hskip -0.3cm \epsfbox{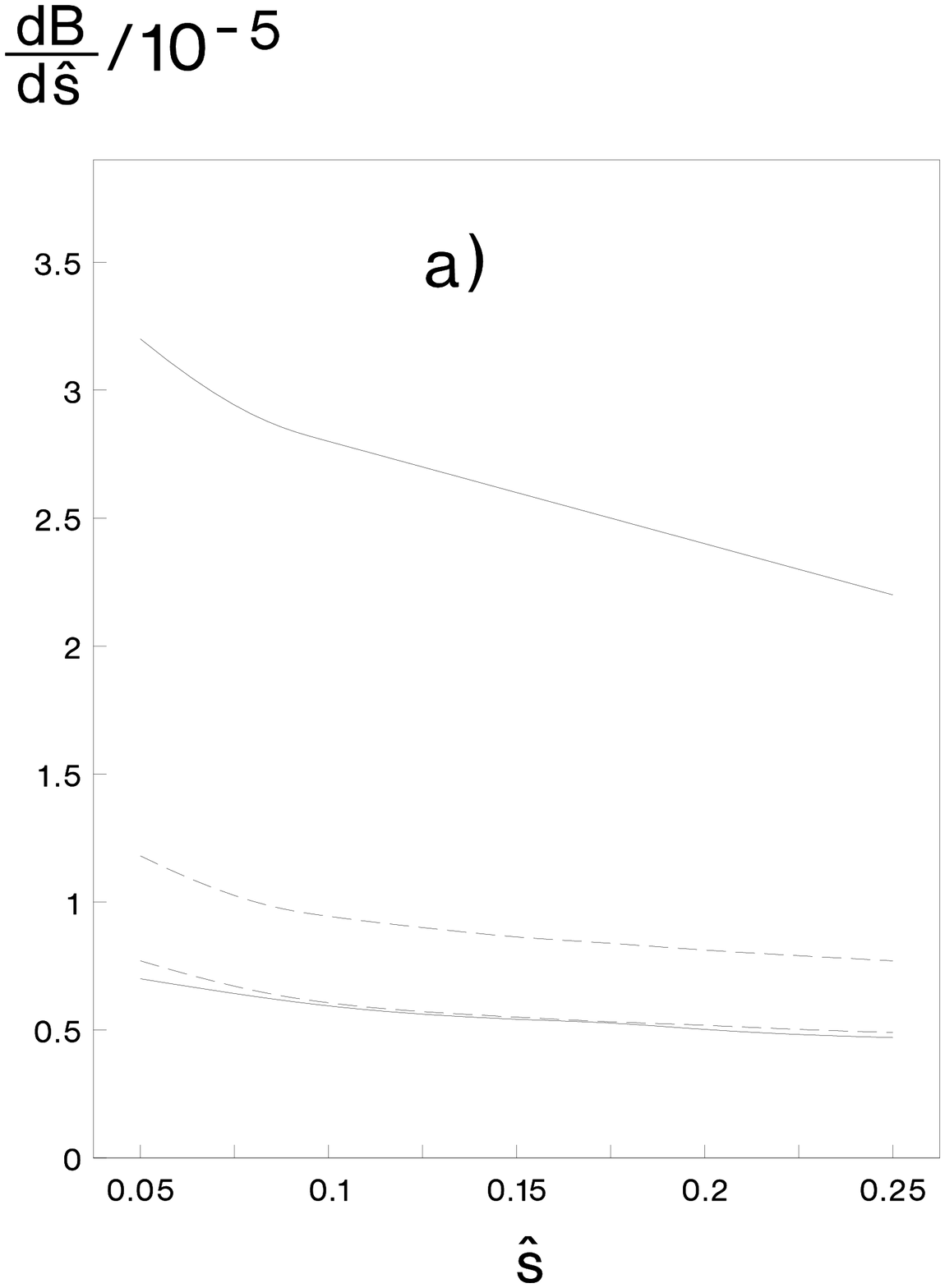}}
\end{flushleft}
\vspace{-13.45cm}
\begin{flushright}
\leavevmode
\epsfysize=12.5cm  
\mbox{\hskip -0.3cm \epsfbox{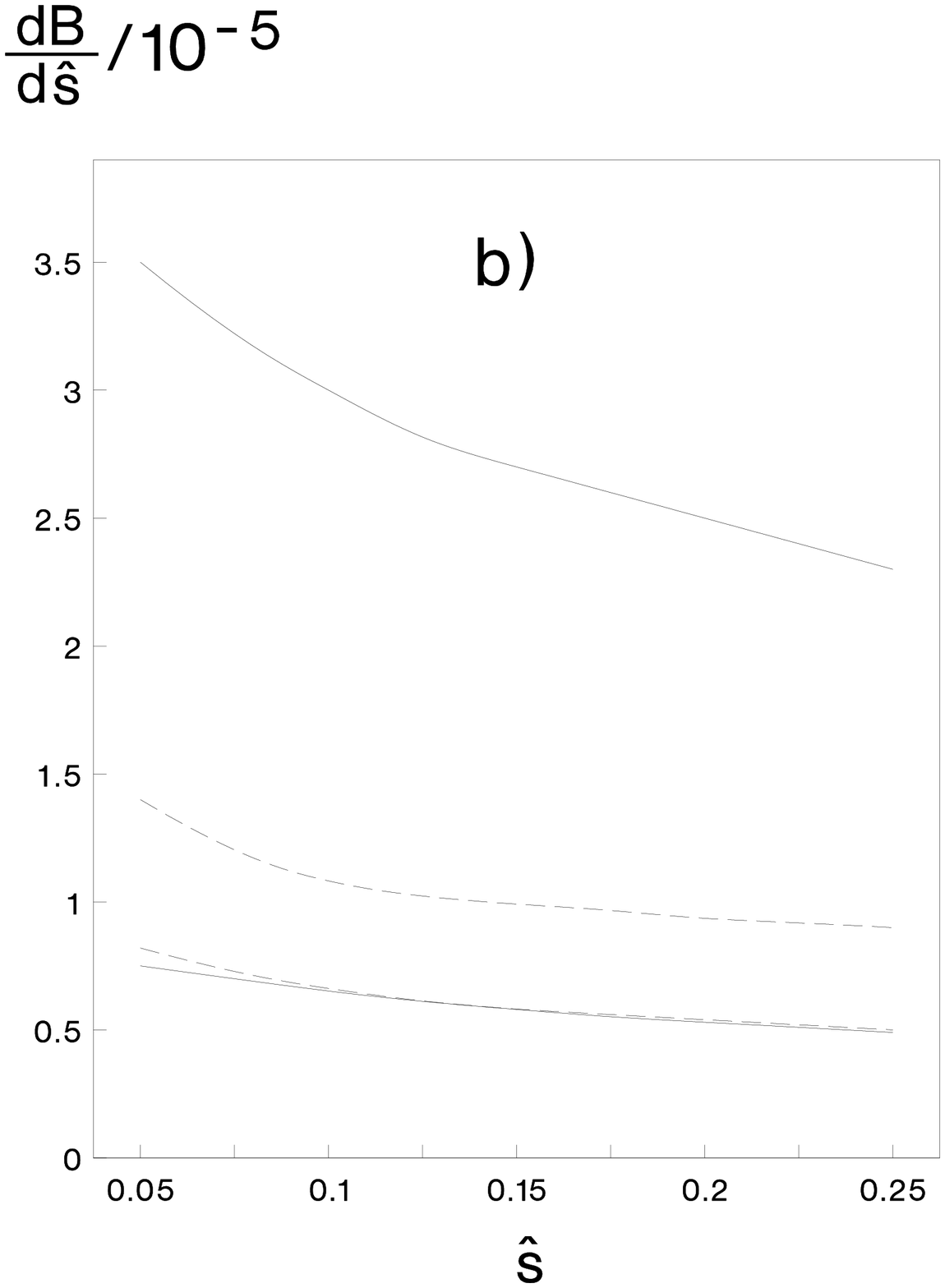}}
\end{flushright}
\ve{-1cm}
\caption{Minimum and maximum values of
$dB(\bxee)/d\hs$ in the SM (dashed lines) and 2HDM (solid lines) a) in the
NNLO, b) in the NLO.} 
\end{figure}

Let us return back to the Model~III. The results derived for the most general
case (when all of three sources are actual) are presented in the Fig.~5. As one
can see from this figure, while the minimum value of $dB(\bxee)/d\hs$ coincides
(within 10\% accuracy) with that in the SM, the maximum value is $2.5 \div
3$ times larger than in the standard model. Note that \oas corrections do not
enhance or suppress the deviation from the SM predictions. This is because in 
both the NLO and the NNLO the experimental constraints on the $\bxg$ 
branching fix
$|C_7^{eff}|$ approximately in the same range (at least for the choice of the
heavy mass scale $\mu_W=m_t$). 
 
For the partially integrated branching ratio
\beq
\Delta B(\bxee)=\int_{0.05}^{0.25}{d\hs\ \frac{dB(\bxee)}{d\hs}} 
\eeq
one gets: \\
SM, NLO: $\Delta B(\bxee)=(1.3 \div 2.1) \times 10^{-6}$ \\
2HDM, NLO: $\Delta B(\bxee)=(1.3 \div 5.6) \times 10^{-6}$ \\
SM, NLO: $\Delta B(\bxee)=(1.2 \div 1.9) \times 10^{-6}$ \\
2HDM, NLO: $\Delta B(\bxee)=(1.1 \div 5.3) \times 10^{-6}$ \\
As it follows from these results, in the 2HDM partially integrated branching
ratio of $\bxee$ decay can be up to 2.8 times larger than in the standard model.
Again, the deviations from the SM result are not affected by \oas corrections.

Thus, if both $\bxg$ and $\bxee$ branching ratios are computed with the same
accuracy, then in any order of the perturbation theory 
(unless in some numerically
relevant order $C_7^{eff}$ becomes highly sensitive to $\hs$) one will derive
that in the 2HDM, maximum value of the
$\bxee$ branching ratio is $2.5 \div 3$ times larger than in
the standard model. \\
\begin{figure}
\ve{-1cm}
\begin{center}
\leavevmode
\epsfysize=12.5cm
\mbox{\hskip -0.7cm \epsfbox{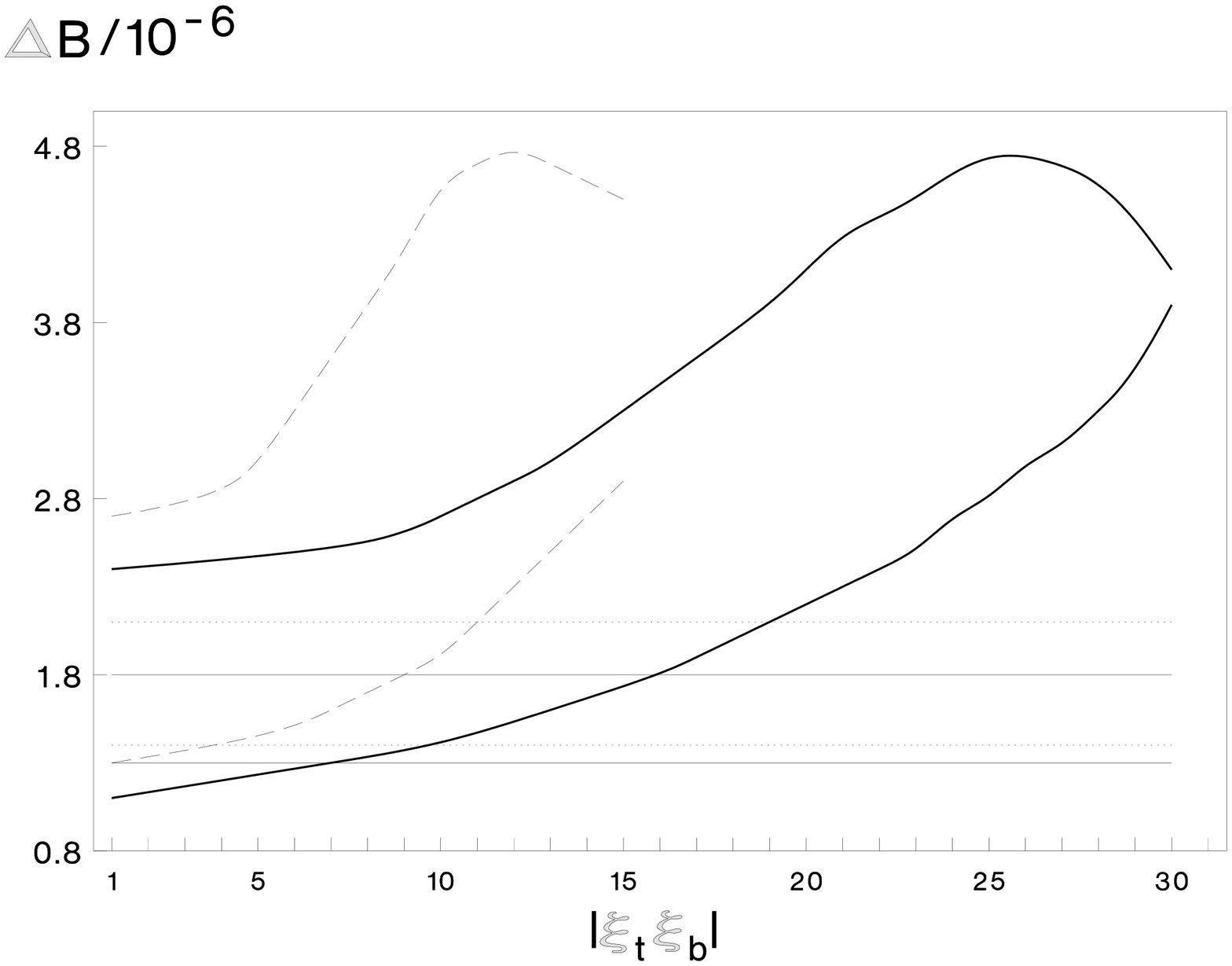}}
\end{center}
\vspace{-0.5cm}
\caption{Minimum and maximum values of the partially integrated 
branching ratio of $B \to X_s e^+ e^-$ decay 
as functions of $|\xi_t \xi_b|$ for $m_{H^+}=400$GeV
in the NLO (dashed lines) and NNLO (fat solid lines). Straight thin solid 
and dotted
lines represent the SM interval respectively in the NNLO and the NLO.}
\end{figure}
\\
8. When considering whole allowed range of the 2HDM parameter space, one can
find out the difference between the NLO and NNLO predictions for the $\bxee$
branching, examining the dependence of this variable on the new physics
parameters. Once in the 2HDM \oas terms are important primarily 
for $C_{7,new}^{eff}$,
it is reasonable to take $m_{H^+}=400GeV$, to minimize the contribution of the
source~III, and investigate the dependence of the $\bxee$ branching ratio 
on the product $|\xi_t \xi_b|$ (see formula (8) and the comment beneath).
The minimum and maximum values of the partially integrated $\bxee$ branching
ratio as functions of $|\xi_t \xi_b|$  are presented in the Fig.~6. One can 
see
that the dependence of $\Delta B$ on $|\xi_t \xi_b|$ is quite different 
in the NLO and the NNLO. Thus, in the next-to-leading order maximum value of 
$\Delta B(\bxee)$ is two and more times larger than in the standard model, when
taking $8 \leq |\xi_t \xi_b| \leq 15$. In the next-to-next-to-leading this
occurs for $17 \leq |\xi_t \xi_b| \leq 30$. In other words, after including \oas
corrections, the dependence of $\Delta B(\bxee)$ on $|\xi_t \xi_b|$ is changed 
{\it drastically}.

Generally speaking, the behavior of $\Delta B(\bxee)$ with $|\xi_t \xi_b|$
can further be modified by higher order effects in the perturbation theory.
On the other hand, it has been shown that in the next-to-next-to-leading order,
when choosing the matching scale as $\mu_W=m_t$, the low-energy scale
dependence of the $\bxee$ decay width is weak. In order to preserve the weak
sensitivity of $R(\hs)$ to $\mu$, higher order corrections to the 
$\bxee$ branching
must somehow cancel each other. This allows one to expect that
higher order effects in the perturbation theory
will not modify the obtained result so drastically, as it is the case
in the Fig.~6. In other words, one may suppose that the dependence of 
$\Delta B(\bxee)$ on $|\xi_t \xi_b|$, derived in the NNLO, is
close enough to the proper one. However, it is necessary to emphasize that such a
suggestion may be done only for specific choices of the heavy mass scale (like
above). More generally (for instance for $\mu_W=M_W$) the NNLO 
results\footnote{When taking $\mu_W=M_W$, one can examine the dependence of the
$\bxee$ branching on the new physics parameters, requiring that 
$0 < |C_{7,new}^{eff}(\mu)|^2_{\hs=0} \leq 0.13$.} for the $\bxee$ branching
can be highly unreliable and be
largely modified by higher order effects in the perturbation theory.
 
Summarizing the discussion of this section one may conclude that including the
\oas effects is an important step on the way of establishing the proper
relations between the $\bxee$ branching and the new physics parameters.  \\
\\
9. Thus, \oas corrections to $\bxee$
decay have been examined in the two-Higgs doublet extension of the standard 
model.
The investigations have been performed for the most general version of the 
2HDM (Model~III) in the so-called off-resonance region of the dilepton invariant
mass ($0.05 < \hs < 0.25$). 

It has been shown that in the case when the new physics effects are sizable, 
$O(\alpha_s)$ corrections (for the fixed values of the 2HDM parameters)
suppress the $\bxee$ decay width 
$1.5 \div 3$ times. 
It is natural to suppose
that $O(\alpha_s^2)$ corrections to the $\bxee$ branching will be numerically 
relevant as well.

Last suggestion is confirmed by the fact that the obtained results are
still sensitive to the choice of the heavy mass scale, when the new physics
effects are sizable. Even after
$O(\alpha_s)$ corrections are included, the uncertainty in the $\bxee$ decay
width connected with the choice of the matching scale reaches 25\%. 

On the other hand, $O(\alpha_s)$ corrections reduce
significantly the low - energy scale dependence of the $\bxee$ branching. 
The $\mu$ - error of the obtained results in the NNLO is $\sim 10\%$ or smaller 
(compared to $\sim 25$\% in the NLO). This means that in the next-to-next-to
leading order the reliability of the predictions for the $\bxee$ branching 
ratio increases.

When using the experimental constraints on the $B \to X_s
\gamma$ branching and calculating $B \to X_s \gamma$ and $B \to X_s
e^+ e^-$ decays with the same accuracy, one will
probably get in all orders of the perturbation theory that
in the 2HDM $\bxee$ branching ratio can be about $2.5 \div 3$ times larger
than in the standard model.  
However only after taking into account \oas corrections and probably those of
higher orders, 
one is able to derive the
proper relations between the new physics parameters and the $\bxee$ branching
ratio.
\\

Author is grateful to H. M. Asatrian and H.H. Asatryan for stimulating
discussions. This work has been partially supported by the SCOPES program.

\end{document}